# Optical emission investigation of laser-produced $MgB_2$ plume expanding in an Ar buffer gas

S. Amoruso,[a] R. Bruzzese, N. Spinelli, R. Velotta, X. Wang,

INFM and Dipartimento di Scienze Fisiche, Via Cintia, I-80126 Napoli (Italy)

C. Ferdeghini

INFM and Dipartimento di Fisica, Via Dodecaneso 33, 16146 Genova (Italy)

## Abstract

Optical emission spectroscopy is used to study the dynamics of the plasma generated by pulsed-laser irradiation of a $MgB_2$ target, both in vacuum and at different Ar buffer gas pressures. The analysis of the time-resolved emission of selected species shows that the Ar background gas strongly influences the plasma dynamics. Above a fixed pressure, plasma propagation into Ar leads to the formation of blast waves causing both a considerable increase of the fraction of excited Mg atoms and a simultaneous reduction of their kinetic flux energy. These results can be particularly useful for optimizing $MgB_2$ thin film deposition processes.

[a] Dipartimento di Ingegneria e Fisica dell'Ambiente, Università della Basilicata C.da Macchia Romana, I-85100 Potenza, Italy. Electronic mail: amoruso@unibas.it

The recent discovery of superconductivity in $MgB_2$ with a transition temperature of ≈ 39 K has attracted much interest due to its rather simple binary intermetallic character and for its potential for future electronics.[1,2] Despite the efforts in producing samples in the form of thin films with different deposition techniques, magnesium-diboride thin film deposition has appeared a quite difficult task due to the high volatility of Mg. There have been reports of $MgB_2$ films prepared by *ex situ* and post-annealing processes and pulsed laser deposition (PLD).[3,4] Nevertheless, the *in situ* deposition process of epitaxial $MgB_2$ film is preferred for the realization of high-tech applications, such as Josephson junctions and heterostructures. In very recent papers both a multilayer PLD technique, involving interposed $MgB_2$ and Mg layers in high vacuum condition,[5] and *in situ* PLD in an Argon pressure[6,7] have been proposed for the growth of superconducting thin films. In the last two reports, a strong effect of the Ar background pressure on plume emission has been observed. In particular, visual inspection of the $MgB_2$ plasma plume showed that the Ar pressure strongly affected the colour of the plume. This indicates that the laser ablation process can be influenced to a great extent by the ambient Ar atmosphere. Thus, from the prospective of controlling $MgB_2$ film growth, it is important to know in detail how the background Ar pressure influences plume chemistry and particles flux.

We report in this letter an optical emission investigation of the plume produced by excimer laser ablation of a $MgB_2$ target both in vacuum and at different Ar buffer pressures. We have carried out time integrated and time- and space-resolved studies of the strongest emission lines of neutral and ionic excited species of the laser ablated plasma over a broad wavelength range (200-800 nm). In particular, our analysis has allowed us to relate the change in the plasma emission, above a specific Ar pressure, to the onset of a blast wave in the plume expansion into the buffer gas. This leads to a considerable increase in the plume excitation (and ionization) and to a reduction of the particles kinetic energies, namely, conditions which can be very helpful for $MgB_2$ thin films deposition by laser ablation.

A XeF excimer laser (351 nm) was used to produce ablation of a stoichiometric $MgB_2$ target.[7] The laser pulse (≈20 ns FWHM) was focused onto the $MgB_2$ target with a resulting energy fluence of ≈3 J $cm^{-2}$. The target was mounted on a rotating holder and placed in a vacuum chamber evacuated to a residual pressure of $10^{-6}$ mbar. During the experiments the chamber was filled with Ar gas whose pressure was varied in the range $10^{-4}$÷1 mbar.

The bright plasma emission was viewed through a side window at right angles to the plume expansion direction. A slice of the plasma was imaged onto the entrance slit of a 0.25 m monochromator equipped with either a 1200 or 100 grooves/mm grating (maximum resolution ≈0.05 nm). The monochromator exit was coupled to either an intensified charged coupled device (ICCD) camera, with a minimum temporal gate of 5 ns and an overall spectral resolution of about 0.4 nm, for time-integrated measurements, or to a photomultiplier tube to trace the temporal evolution of selected spectral lines of Mg and B excited atoms and ions.

Figure 1 reports time integrated emission spectra from the $MgB_2$ expanding plume at different Ar background pressures for a representative distance $d$=5 mm from the target surface. While the arrival of the laser pulse on target provided a built-in zero time, the integration time was chosen as to collect all the light emitted for a given Ar pressure (typically 20-50 μs). In these measurements the monochromator, coupled to the ICCD camera, was equipped with a 100 grooves/mm grating, in order to register simultaneously all the emission lines in the broadest possible wavelength range. The 1200 grooves/mm grating was then used to obtain a more reliable identification of the most significant lines in the investigated spectral range. The identification of Mg and B lines was accomplished by standard data available in the literature.[8] The broad peak in the UV (around 285 nm) in Fig. 1 is due to the overlapping of spectrally very close emission lines of excited B and Mg atoms and ions, unresolved by the 100 grooves/mm grating.

The spectra mainly consist of neutral B and Mg lines (B I and Mg I), with some contributions from their excited ions (B II and Mg II), and the plume emission in the visible range can be mainly ascribed to Mg I and Mg II, since the most intense B lines belong to the UV spectral range. In particular, the interesting feature is that while at low buffer pressure (up to ≈$10^{-1}$ mbar) the contribution from the Mg I 518 nm (3s4s $^3S$ – 3s3p $^3P^0$) green line to the emitted visible light is larger then that of the Mg II blue line at 448 nm (4f $^2F^0$ – 3d $^2D$), this last line strongly increases as the pressure grows up, becoming almost an order of magnitude larger than the green line at the larger pressures.

The presence of Ar leads to a variation in the plume colour and shape, as reported in refs. 6 and 7 ; in particular, in our experimental conditions, a green plume was produced close to the target for an Ar pressure up to ≈$7\times10^{-2}$ mbar. By increasing the background gas pressure above this value a long and bright sky blue plasma plume was observed, the plume resulting more confined for larger pressures (up to 1 mbar). Then, the difference in the plume emission observed at different Ar background pressures can be totally ascribed to the excitation and ionisation dynamics of the different species contributing to the visible plume.

In Figure 2 the time-integrated photon yields of three intense Mg excited atom and ion lines of the spectrum of Fig. 1 are reported as a function of the Ar background pressure at a distance *d*=5 mm from the target. The general trend, also observed for other emission lines of Fig. 1, is that of a strong decrease of the photon yield up to an Ar pressure of ≈$7\times10^{-2}$ mbar, followed by a steep increase at higher pressures. This evidences the central role played by the Ar buffer gas in determining the plasma kinetics by promoting excitation and ionization of the plume species and, in particular, of Mg atoms.

Time resolved measurements of different emission lines were also performed to elucidate plume dynamics. Figure 3 shows the time-of-flight (TOF) distribution of a Mg I emission line ($\lambda$=383 nm corresponding to the transition 3s3d $^3D$ – 3s3p $^3P^0$), at a distance of 5 mm and at different pressures. At background pressure of $2\times10^{-1}$ mbar the signal was long lasting, indicating the occurrence of a slowing down effect due to collisions between the Mg and Ar atoms. The inset of Fig. 3 shows the dependence of the time $t_p$ at which the peak in the TOF distribution is observed on the distance $d$ from the target. The results obtained at pressures smaller than $10^{-1}$ mbar show a linear dependence over the whole range of distances for which emission could be detected. On the contrary, when the Ar pressures exceeds $10^{-1}$ mbar, the dependence is no longer linear for $d$ larger than 3 mm. Similar features were observed for other emission lines. The velocities calculated from the slopes of the $d$-$t_p$ curves in the linear region are of the order of ≈ $10^6$ cm/s, while at the larger pressure the asymptotic velocity is of the order of ≈ $2\times10^4$ cm/s. This clearly shows that the Ar gas environment strongly affects plasma expansion dynamics for pressures of ≈$10^{-1}$ mbar at distances $d$ of few mm from the target surface.

The interaction of the plume with a background gas is usually modelled as a scattering of ablated species due to elastic collisions with the gas atoms.[9] In this respect, it is worth reporting that at the gas pressures used in this experiment the average value of the mean free path of a Mg atom with kinetic energies of tens of eV in an Ar background scales from ≈ 20 mm at $10^{-2}$ mbar and to ≈2 mm at $10^{-1}$ mbar, respectively, as estimated by using the equations reported by Westwood.[10] Thus, several collisions occur between ejected species and buffer gas atoms for distances of different mm from the target at an Ar pressure threshold of ≈$10^{-1}$ mbar. Moreover, due to the inverse dependence of the mean free path on the buffer gas pressure, a slowing down effect will occur at larger distances from target for lower Ar buffer pressures. Notwithstanding, simulation on Si (whose mass and radius are close to that of Mg) plume expansion in Ar gas have shown that slowing down and plume splitting occurs after just few collisions at $10^{-1}$ mbar leading to the generation of a blast wave.[9] Our data (see inset of Fig. 3) are well described by a blast wave for $d$>3 mm at an Ar pressure of $2\times10^{-1}$ mbar.

Blast waves have been largely considered to describe laser ablated plumes expansion into ambient gas. Recently, a simplified model describing the expansion of the plume into an ambient gas has been reported and compared to experimental data.[11] The model shows that with the formation of a shock wave a redistribution of kinetic and thermal energies occur between the plume and the ambient gas, leading to a significant transformation of plume stream velocity into thermal energy.

To support this interpretation, the Boltzmann plot method was used to evaluate the temperature of Mg I and to relate this to the electron temperature. A typical Boltzmann plot obtained at a pressure of $2\times10^{-2}$ is reported in the inset of Fig. 4 which shows the dependence of the mean electron temperature $T_e$ on the Ar buffer pressure $p$ again at the representative value $d$=5 mm. The data clearly show a slight decrease of $T_e$ for $p<10^{-1}$ mbar followed by a steep increase above $10^{-1}$ mbar, marking a turnover between two different regimes around $7\times10^{-2}$ mbar. The minimum of $T_e$ can be explained by considering that the plume exchanges energy with the buffer gas through collisions as soon as the mean free path becomes comparable with the distance of the experiment. Then by increasing the buffer pressure, a larger number of collisions occur, compressing the ambient gas to an extent such that a shock wave develops, and the transformation of plume kinetic energy into thermal energy takes place. Then, the heating of the plume shown in Fig. 4 leads to an enhancement of plume excitation and ionisation at the expense of the particle kinetic energy, as observed above.

In summary, optical emission spectroscopy has allowed us to observe that when a laser ablated $MgB_2$ plume expands in an Ar environment, the fraction of excited Mg atoms is increased while their flux velocity is reduced above a threshold pressure of $10^{-1}$ mbar at few mm from the target surface. This is due to the formation of a blast wave at a distance from the target surface of the order of few mean free paths for a given Ar pressure. Since the excitation state and the kinetic energy of an impinging atom greatly influence the process of thin films formation,[12] the mechanism we have observed lends itself as a powerful means for the control of $MgB_2$ film growth. In fact energetic particles cause resputtering and largely influence the deposit content especially of the most volatile elements, like Mg in $MgB_2$. Thus, a decrease in the energy of the impinging species could contribute to the increase of the Mg content in the deposit. Moreover, the arrival of a larger extent of excited species on the growing film produces an increase of the energy released in the impact region, thus enhancing surface atomic mobility and favouring film growth at lower substrate temperatures. Since the mean free path is inversely proportional to the gas pressure, both the steep increase of the excited species and the reduction of particles velocity, will take place at larger distances from target for lower Ar buffer pressures. As a consequence, there is a "best" buffer gas pressure, as a function of the target-to-substrate distance, which optimizes the amount of Mg plume atoms with a higher degree of internal excitation energy and a lower kinetic energy arriving on a substrate. This conclusion, which summerizes the results of our analysis, is in very good agreement with the recent findings of Ref. 7, where deposition of as grown $MgB_2$ superconducting thin films was successfully obtained only by increasing the Ar buffer pressure up to $5\times10^{-2}$ mbar with a target-to-substrate distance of ≈15 mm. These buffer gas and target-to-substrate distance values perfectly match the conditions indicated by our results for the onset of blast waves produced by the expanding plume in the buffer gas.

## Figure captions

Figure 1: Time-integrated optical emission spectra of the $MgB_2$ plume at different Ar buffer pressures, for $d$=5 mm.

Figure 2: Time-integrated photon yields as a function of the Ar background pressure for Mg I and Mg II emission lines at $d$=5 mm.

Figure 3: TOFD of Mg atoms as a function of the distance at different Ar pressures. Inset: distance $d$ from the target on the time $t_p$ at which the peak in the TOF distribution is observed.

Figure 4: Electron temperature $T_e$ as a function of the Ar ambient pressure. Inset: Boltzmann plot of Mg I emission lines at $d$=5 mm and $p=2\times10^{-2}$ mbar.

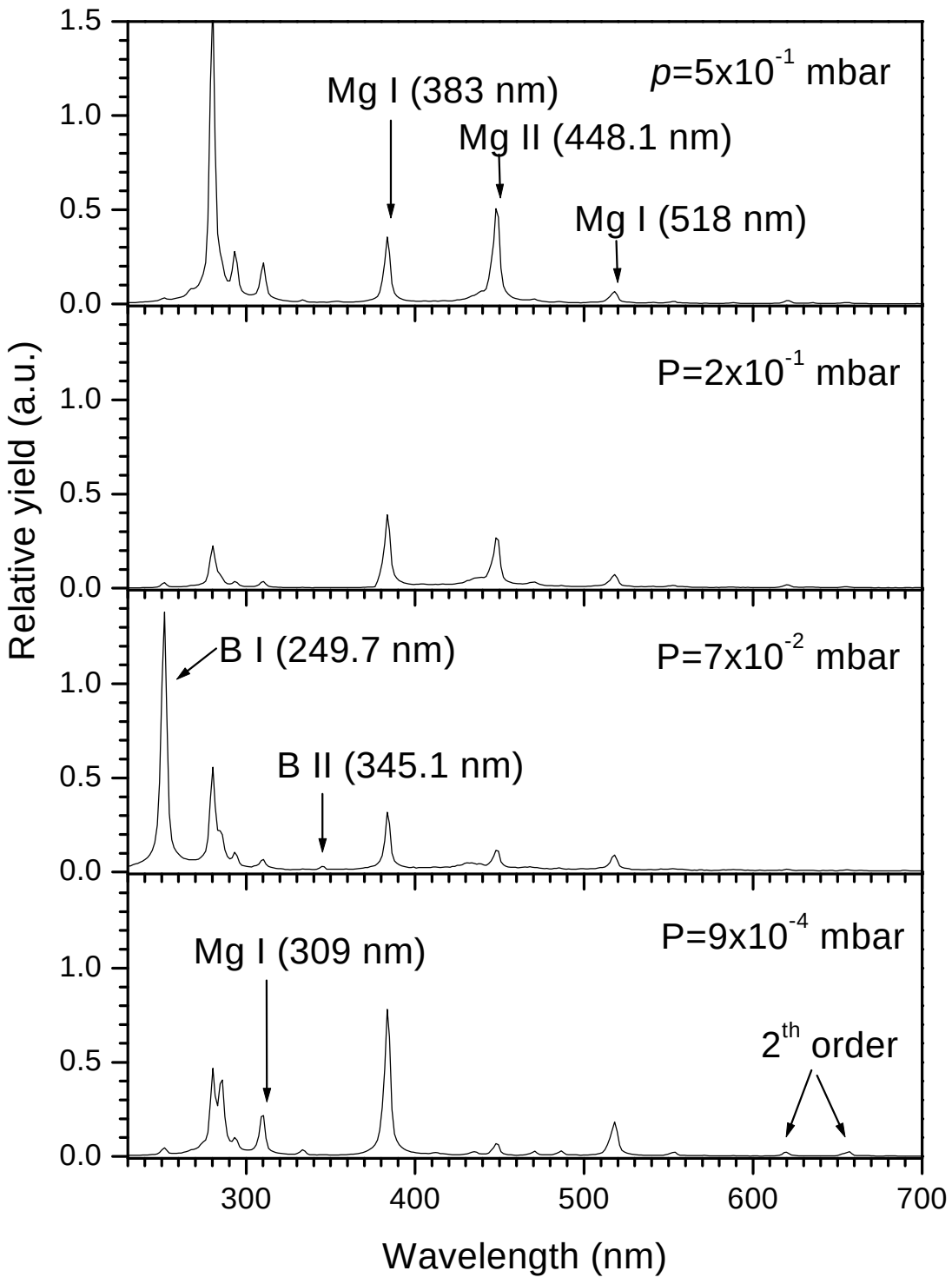

p=5x10^-1 mbar
Mg I (383 nm)
Mg II (448.1 nm)
Mg I (518 nm)
P=2x10^-1 mbar
P=7x10^-2 mbar
B I (249.7 nm)
B II (345.1 nm)
P=9x10^-4 mbar
Mg I (309 nm)
2^th order
Relative yield (a.u.)
Wavelength (nm)

Fig 1

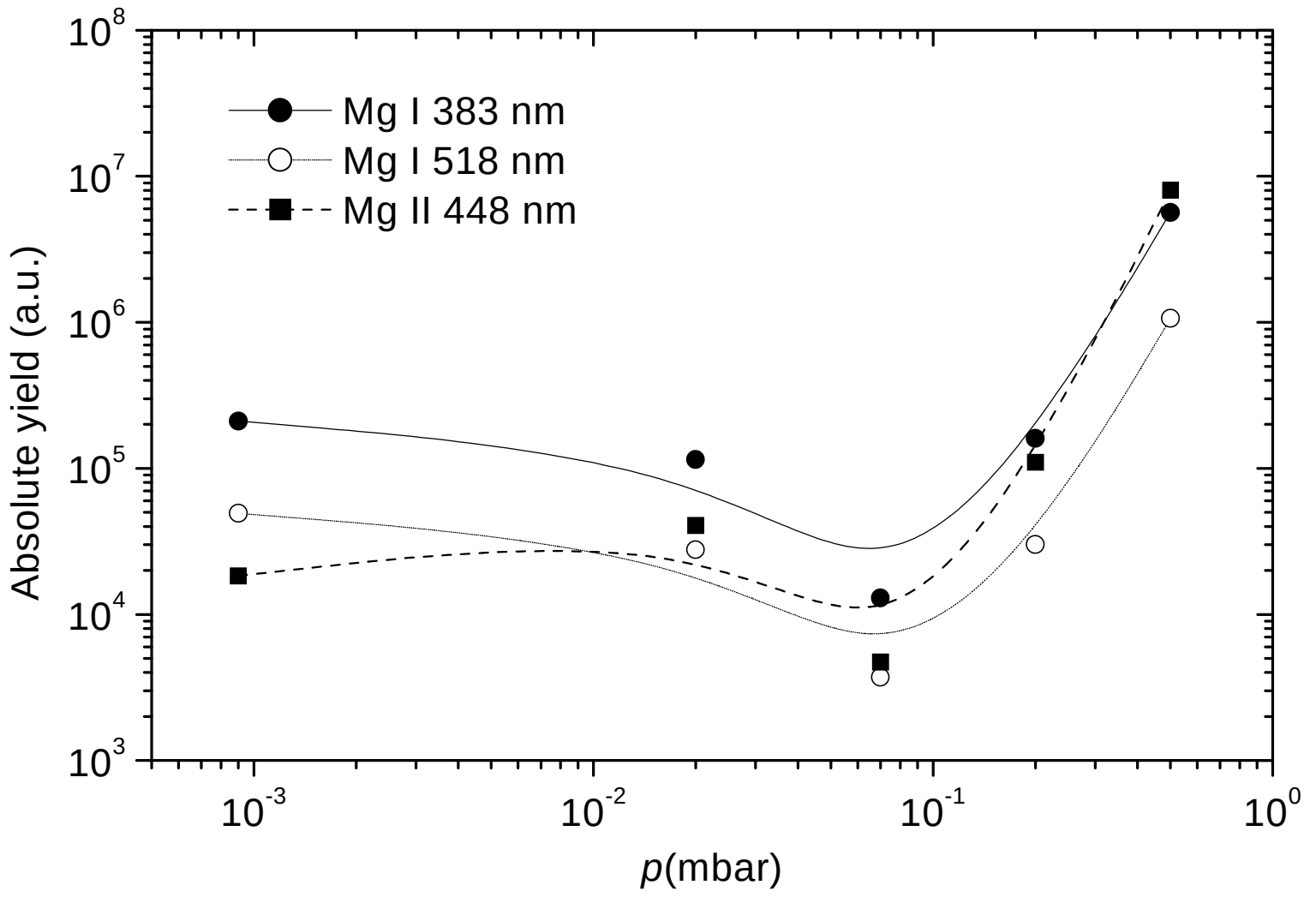

Mg I 383 nm
Mg I 518 nm
Mg II 448 nm
Absolute yield (a.u.)
p(mbar)

Fig.2

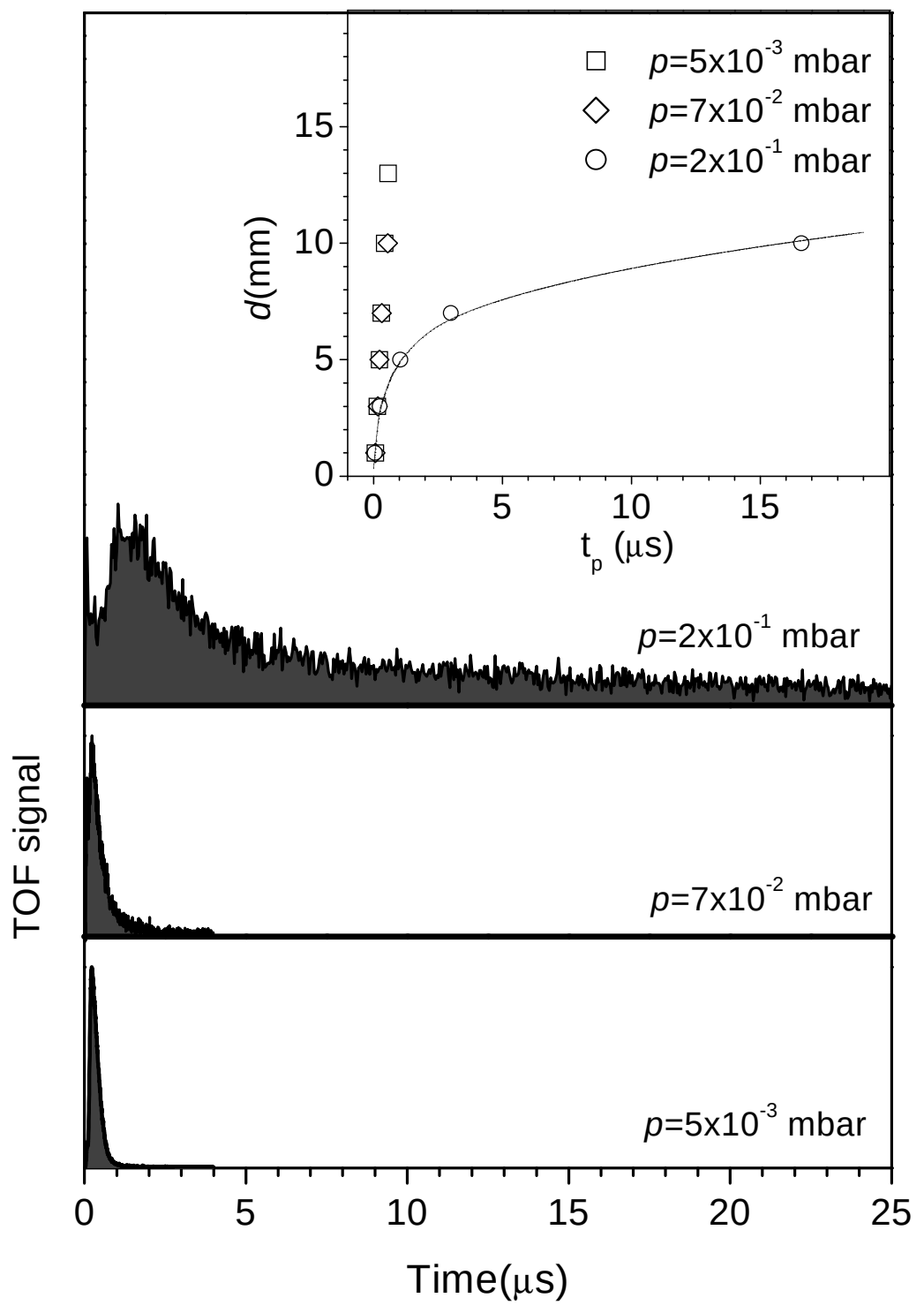

Fig.3

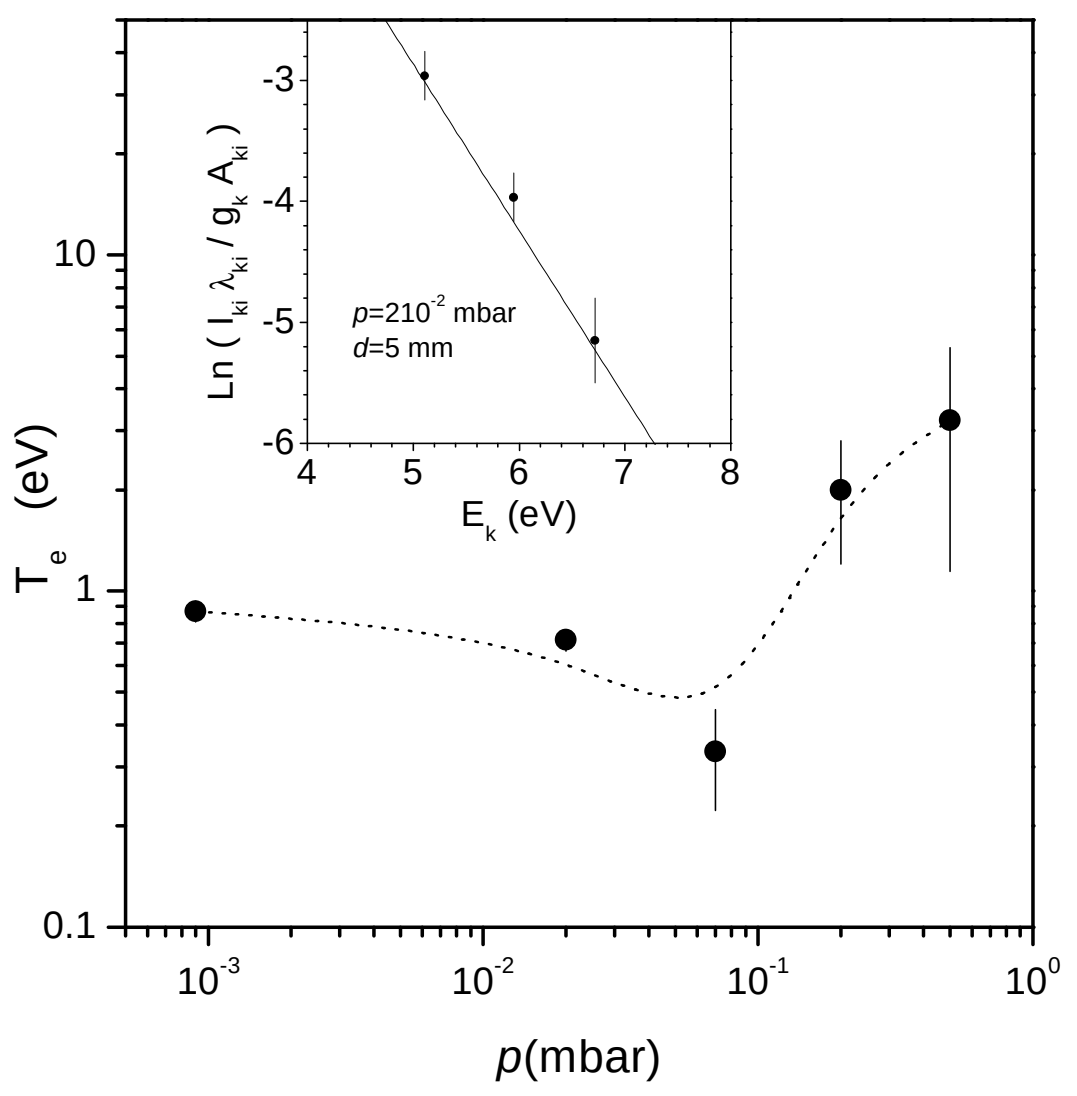

Ln ( Iki λki / gk Aki )
p=210-2 mbar
d=5 mm
Ek (eV)
Te (eV)
p(mbar)

Fig. 4